\documentclass[12pt]{article}
\usepackage{graphicx}
\usepackage[cp1251]{inputenc}

\begin{document}
 \noindent {\footnotesize\it Astronomy Letters, 2018, Vol. 44, No 3, pp. 193--201.}
 \newcommand{\dif}{\textrm{d}}

 \noindent
 \begin{tabular}{llllllllllllllllllllllllllllllllllllllllllllll}
 & & & & & & & & & & & & & & & & & & & & & & & & & & & & & & & & & & & & & \\\hline\hline
 \end{tabular}

  \vskip 0.5cm
 \centerline{\bf\Large Search for Stellar Streams Based on Data from}
 \centerline{\bf\Large the RAVE5 and Gaia TGAS Catalogues}
 \bigskip
 \bigskip
  \centerline
 {
 A.T. Bajkova\footnote [1]{e-mail: anisabajkova@gao.spb.ru} and
 V.V. Bobylev}
 \bigskip

  \centerline{\small\it
 Pulkovo Astronomical Observatory, Russian Academy of Sciences,}

  \centerline{\small\it
 Pulkovskoe sh. 65, St. Petersburg, 196140 Russia}
 \bigskip
 \bigskip
 \bigskip

 {
{\bf Abstract}---We have analyzed the space velocities of stars
with the proper motions and trigonometric parallaxes from the Gaia
TGAS catalogue in combination with the line-of-sight velocities
from the RAVE5 catalogue. In the $V,\sqrt{U^2+2V^2}$ velocity
plane we have identified three clumps, BB17-1, BB17-2, and BB17-3,
in the region of large velocities ($V<-150$~km s$^{-1}$). The
stars of the BB17-1 and BB17-2 clumps are associated with the
kinematic groups VelHel-6 and VelHel-7 detected previously by
Helmi et al. We give the greatest attention to the BB17-3 clump.
The latter is shown to be most closely linked with the debris of
the globular cluster $\omega$~Cen. In the BB17-3 clump we have
identified 28 stars with a low velocity dispersion with respect to
the center of their distribution. All these stars have very close
individual age estimates: $\log t\approx10$. The distribution of
metallicity indices in this sample is typical for the stars of the
globular cluster $\omega$~Cen. In our opinion, the BB17-3 clump
can be described as a homogeneous stream in the debris of the
cluster $\omega$~Cen.
  }

\medskip
DOI: 10.1134/S1063773718020019

 \subsection*{INTRODUCTION}
A study of the ellipsoidal distribution of stellar space
velocities as a function of stellar age led to an understanding of
the dynamical properties of the Galaxy and its subsystems. The
observed velocity distribution turned out to be nonuniform (Eggen
1958, 1996).

The fine structure of the velocity field manifests itself in
different planes. An analysis of the rectangular $U,V,W$
velocities is most popular. Even in the region of relatively small
(up to $\sim$40~km s$^{-1}$) $U$ and $V$ velocities there are
known peaks named after their association with open star clusters
(OSCs), such as the Pleiades, Sirius, Coma Berenices, or the
Hyades (Eggen 1958, 1996). It was suggested to associate a number
of peaks with older open star clusters and stellar groups in the
Galactic disk, for example, Wolf 630, 61~Cygni (Eggen 1969),
$\zeta$~Herculis, $\sigma$~Puppis, or $\varepsilon$~Indi (Eggen
1971a).

According to the theory of stellar streams (Eggen 1996), the peaks
reflect the presence of tubes or stellar debris, OSC debris, in
circumsolar space. Such debris must be formed during the
disruption of clusters under the influence of Galactic tides,
giant molecular clouds, and other gravitational perturbations. The
debris gradually extend along the Galactic orbit of clusters and
are completely mixed with the stellar Galactic background 1--2 Gyr
later. The presence of debris is confirmed by numerical
simulations of the dynamical evolution of OSCs (Chumak et al.
2005; Chumak and Rastorguev and 2006a, 2006b).

An analysis of the stellar velocities from the Hipparcos (1997)
catalogue showed that, in addition to the already known peaks, the
two-dimensional distribution of $UV$ velocities has a more complex
structure. Specifically, there are several extended ``branches''
located almost parallel to one another (Skuljan et al. 1999).

Apart from the branches, bimodality of the distribution manifests
itself in the $UV$ velocity plane (Dehnen 1998, 2000). Such a
velocity stratification can be caused by resonance effects. In
particular, the formation of the $\zeta$~Herculis
($V\approx-50$~km s$^{-1}$) and Wolf 630 ($V\approx-20$~km
s$^{-1}$, $U\approx40$~km s$^{-1}$) streams can be explained by
the resonances produced by the central Galactic bar (Dehnen 2000;
Fux 2001).

Tidal tails have been detected in a number of globular clusters
(Grillmair et al. 1995; Leon et al. 2000; Odenkirchen et al. 2001;
Belokurov et al. 2006; Sollima et al. 2011; Navarrete et al. 2017)
and dwarf galaxies, satellites of the Milky Way Galaxy (Ibata et
al. 2001; Grillmair and Dionatos 2006; Helmi 2008). Some of the
observed clumps of stellar $UV$ velocities in the solar
neighborhood at velocities $V<-100$~km s$^{-1}$ are associated
with such debris. For example, these include the Arcturus (Eggen
1971b) and KFR08 (Klement et al. 2008) streams. It is suggested
that several stellar streams with high velocities $V<-150$~km
s$^{-1}$ consisting of extremely metal-poor stars recently
discovered by Helmi et al. (2017) are the debris of disrupted
satellite galaxies. In the opinion of these authors, most of the
detected streams are associated with the debris of the globular
cluster $\omega$~Cen.

The goal of this study is to analyze the stellar velocity field in
the solar neighborhood to reveal stellar groups with a common
dynamical origin. We focus our attention on the stars with large
velocities relative to the Sun and, as a rule, these are fairly
old, metalpoor stars. For our analysis we use the positions,
proper motions, and trigonometric parallaxes from the Gaia DR1
catalogue (Gaia Collaboration, Prusti et al. 2016) in combination
with the line-of-sight velocities from the RAVE (RAdial Velocity
Experiment) catalogue (Steinmetz et al. 2006).

 \subsection*{DATA}
The observations within the RAVE project have been performed since
2003 in the Southern Hemisphere at the 1.2-m Schmidt telescope of
the Anglo-Australian Observatory. Five releases of the catalogue
have been published. The mean line-of-sight velocity error is
about 3~km s$^{-1}$. The RAVE5 version (Kunder et al. 2017)
contains data on 457 588 stars; more than 200 000 of them are
common to the first published version of the Gaia catalogue.

This version was produced from a combination of the data in the
first year of orbital Gaia satellite observations with the Tycho-2
stellar positions (Hog et al. 2000). It is designated as TGAS
(Tycho.Gaia Astrometric Solution; Brown et al. 2016; Lindegren et
al. 2016) and contains the positions, trigonometric parallaxes,
and proper motions of $\sim$2 million stars. For $\sim$90 000
stars common to the Hipparcos (1997) catalogue the mean random
error of their proper motions is $\sim$0.06 mas yr$^{-1}$
(milliarcseconds per year); for the remaining stars this error is
$\sim$1 mas yr$^{-1}$ (Brown et al. 2016).

 \subsection*{METHOD}
We know three stellar velocity components from observations: the
line-of-sight velocity $V_r$ and the two velocity components
$V_l=4.74r\mu_l \cos b$ and $V_b=4.74r\mu_b$ along the Galactic
longitude $l$ and latitude $b$ expressed in km s$^{-1}$. The
proper motion components $\mu_l \cos b$ and $\mu_b$ are expressed
in mas yr$^{-1}$. The coefficient 4.74 is the ratio of the number
of kilometers in an astronomical unit to the number of seconds in
a tropical year, and $r$ is the stellar heliocentric distance in
kpc. In this paper the distances are calculated using the
trigonometric parallaxes from the Gaia TGAS catalogue and,
therefore, $r=1/\pi.$ The $U,V,W$ velocities along the rectangular
Galactic $x,y,z$ coordinate axes are calculated via the $V_r, V_l,
V_b$ components:
 \begin{equation}
 \begin{array}{lll}
 U=V_r\cos l\cos b-V_l\sin l-V_b\cos l\sin b,\\
 V=V_r\sin l\cos b+V_l\cos l-V_b\sin l\sin b,\\
 W=V_r\sin b                +V_b\cos b.
 \label{UVW}
 \end{array}
 \end{equation}
where the velocity $U$ is directed from the Sun to the Galactic
center, $V$ is in the direction of Galactic rotation, and $W$ is
directed to the North Galactic Pole. To extract the statistically
significant signals of nonuniformities in the distributions of
$UV$ velocities, we use the wavelet transform known as a powerful
tool for filtrating spatially localized signals (Chui 1997;
Vityazev 2001). The wavelet transform of the two dimensional
distribution $f(U,V)$ consists in its decomposition into analyzing
wavelets $\psi(U/a,V/a)$, where $a$ is the coefficient that allows
the wavelet of a certain scale to be extracted from the entire
family of wavelets characterized by the same shape $\psi$. The
wavelet transform $w(\xi,\eta)$ is defined as a correlation
function in such a way that at any given point $(\xi,\eta)$ in the
$UV$ plane we have one real value of the following integral:
 \begin{equation}
 \renewcommand{\arraystretch}{2.2}
 \begin{array}{ccc}\displaystyle
 w(\xi,\eta)=\int_{-\infty}^\infty \int_{-\infty}^\infty f(U,V)
   \psi\Biggl(\displaystyle\frac{(U-\xi)}{a},\frac{(V-\eta)}{a}\Biggr) dU dV,
 \end{array}
 \end{equation}
which was called the wavelet coefficient at point $(\xi,\eta)$.
Obviously, in the case of finite discreet maps that we deal with,
their number is finite and equal to the number of square bins on
the map.

%%%%%%%%%%%%%%%%%%%%%%%%%%%%%%%%%%%%%%%%% Fig1
 \begin{figure}[t]{\begin{center}
 \includegraphics[width=100mm]{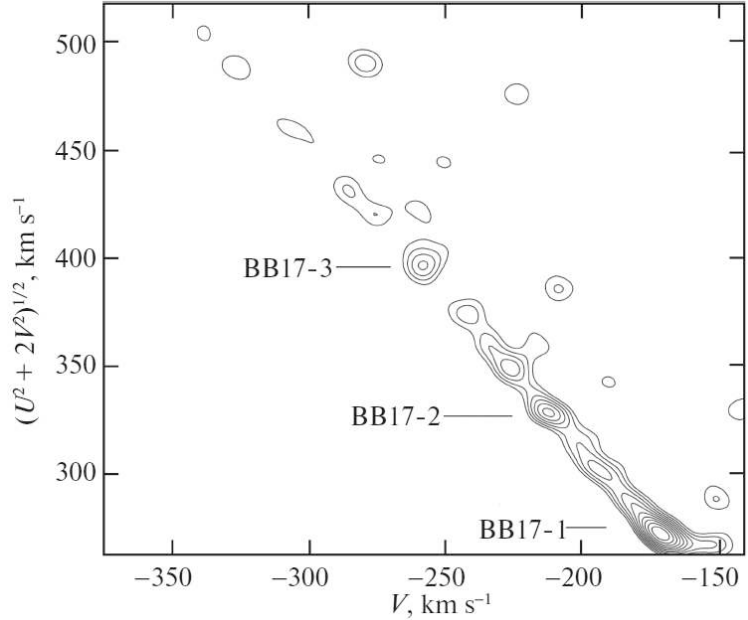}
 \caption{
Wavelet map of the $V,\sqrt{U^2+2V^2}$ velocity plane. The stellar
velocities are given relative to the Sun.
 }
 \label{f-2} \end{center} } \end{figure}
%%%%%%%%%%%%%%%%%%%%%%%%%%%%%%%%%%%%%%%%%%%%%%%%%%%%%%%

As the analyzing wavelet we use a traditional wavelet called the
Mexican HAT (MHAT). The two-dimensional MHAT wavelet is described
by the expression
\begin{equation}
 \renewcommand{\arraystretch}{1.2}
 \psi(d/a)=\Biggl(2-\frac{d^2}{a^2}\Biggr)e^{-d^2/2a^2},
 \label{w4}
 \end{equation}
where $d^2=U^2+V^2$. The wavelet (3) is obtained from
differentiating the Gaussian function twice. The main property of
the function $\psi$ is that its integral over $U$ and $V$ is zero,
which allows any nonuniformities in the distribution being studied
to be detected. If the distribution being analyzed is uniform,
then all coefficients of the wavelet transform will be zero.

To select the candidates without significant random observational
errors, we took the stars satisfying the following criteria:
\begin{equation}
 \begin{array}{ccc}
 |V_r|<350~{\hbox {\rm km s$^{-1}$}},\\
 \sigma_{V_r}<5~{\hbox {\rm km s$^{-1}$}},\\
 \sigma_{\mu_\alpha\cos\delta}<10~{\hbox {\rm mas yr$^{-1}$}},\\
 ~~~~\sigma_{\mu_\delta}<10~{\hbox {\rm mas yr$^{-1}$}},\\
 \sigma_\pi/\pi<0.5.
 \end{array}
 \end{equation}
More than 200 000 stars with the trigonometric parallaxes and
proper motions from the Gaia TGAS catalogue and the line-of-sight
velocities from the RAVE5 catalogue satisfy these constraints. The
constraint on the magnitude of the line-of-sight velocity $|V_r|$
is related to the existence of an appreciable number of
low-quality measurements (with low signal-to noise ratios) with
line-of-sight velocities exceeding the threshold escape velocity
from the Galaxy.

%%%%%%%%%%%%%%%%%%%%%%%%%%%%%%%%%%%%%%%%%%%%%%%%%%%%%%% table~1.
 \begin{table}[t]
 \caption[]{\small
 Data on the stars of the BB17-3 clump
 }
  \begin{center}  \label{t:01}
  \small
  \begin{tabular}{|l|r|r|r|r|r|r|r|}\hline
 ~~~ TYC      & $l,$ deg. & $b,$ deg. & $T_{\hbox {\tiny\rm eff}},$ K & $\log~${\it g} & [M/H], dex & $\pi_{Gaia},$ mas & $\log t$~~~~ \\\hline
 8004-1288-1  & $   1.4$ & $ -62.5$ & 5212 & 2.94 & $-1.83\pm0.18$ & $0.67\pm0.33$ & $10.00\pm0.05$ \\
 7954-0027-1  & $   1.8$ & $ -36.7$ & 5076 & 2.76 & $-1.86\pm0.18$ & $1.07\pm0.32$ & $ 9.99\pm0.07$ \\
 6989-0218-1  & $   4.8$ & $ -78.1$ & 4993 & 2.72 & $-0.85\pm0.09$ & $1.95\pm0.41$ & $ 9.89\pm0.22$ \\
 7442-1834-1  & $  11.9$ & $ -32.0$ & 5801 & 0.78 & $-0.47\pm0.09$ & $5.35\pm0.29$ & $            $ \\
 6900-0414-1  & $  14.8$ & $ -27.1$ & 5725 & 3.98 & $-2.12\pm0.26$ & $1.26\pm0.29$ & $ 9.98\pm0.12$ \\
 6410-0404-1  & $  48.2$ & $ -73.1$ & 4931 & 2.13 & $-1.59\pm0.14$ & $1.52\pm0.34$ & $ 9.99\pm0.07$ \\
 4696-0131-1  & $ 171.8$ & $ -62.2$ & 5931 & 3.87 & $-1.06\pm0.19$ & $3.53\pm0.30$ & $ 9.93\pm0.20$ \\
 7028-0807-1  & $ 229.0$ & $ -52.5$ & 5457 & 2.73 & $-2.35\pm0.27$ & $0.86\pm0.28$ & $10.01\pm0.06$ \\
 7042-1158-1  & $ 236.3$ & $ -41.5$ & 4962 & 1.86 & $-1.36\pm0.17$ & $1.61\pm0.25$ & $ 9.99\pm0.11$ \\
 7018-0652-1  & $ 238.2$ & $ -62.0$ & 5109 & 3.60 & $-1.01\pm0.19$ & $2.58\pm0.23$ & $ 9.91\pm0.19$ \\
 6520-0236-1  & $ 239.0$ & $ -15.0$ & 4985 & 1.87 & $-1.33\pm0.17$ & $0.94\pm0.31$ & $ 9.88\pm0.28$ \\
 8075-0797-1  & $ 257.5$ & $ -43.9$ & 4031 & 0.55 & $-1.23\pm0.14$ & $0.62\pm0.27$ & $10.00\pm0.08$ \\
8890-0860-1 * & $ 276.5$ & $ -34.1$ & 3800 & 1.00 & $ 0.00\pm0.10$ & $1.25\pm0.45$ & $            $ \\%===
 8048-0222-1  & $ 277.4$ & $ -62.3$ & 4655 & 0.82 & $-2.03\pm0.14$ & $1.02\pm0.32$ & $10.00\pm0.05$ \\
 6658-1026-1  & $ 279.8$ & $  32.2$ & 5434 & 3.05 & $-1.31\pm0.18$ & $4.31\pm0.43$ & $            $ \\
 6656-0054-1  & $ 282.0$ & $  34.9$ & 5500 & 3.77 & $-2.15\pm0.26$ & $3.54\pm0.36$ & $10.00\pm0.08$ \\
 9152-1687-1  & $ 285.6$ & $ -41.5$ & 5659 & 4.74 & $-2.10\pm0.26$ & $2.80\pm0.29$ & $10.00\pm0.10$ \\
 8227-1820-1  & $ 291.6$ & $   9.2$ & 4794 & 0.80 & $-2.82\pm0.14$ & $0.70\pm0.26$ & $10.00\pm0.06$ \\
 4954-0606-1  & $ 296.1$ & $  56.3$ & 5038 & 1.74 & $-1.54\pm0.17$ & $1.30\pm0.49$ & $10.00\pm0.12$ \\
 8852-0851-1  & $ 298.1$ & $ -53.8$ & 5055 & 4.39 & $-2.04\pm0.26$ & $0.77\pm0.25$ & $ 9.99\pm0.10$ \\
 8473-0759-1  & $ 300.4$ & $ -63.4$ & 4836 & 2.44 & $-0.62\pm0.09$ & $1.13\pm0.24$ & $ 9.76\pm0.25$ \\
 6112-0045-1  & $ 308.7$ & $  46.9$ & 5846 & 3.56 & $-2.33\pm0.26$ & $2.59\pm0.92$ & $10.00\pm0.07$ \\
 9032-2385-1  & $ 314.6$ & $  -6.3$ & 4921 & 1.97 & $-1.33\pm0.17$ & $1.36\pm0.34$ & $ 9.78\pm0.40$ \\
 8821-0626-1  & $ 333.3$ & $ -48.3$ & 5750 & 3.21 & $-1.32\pm0.18$ & $0.62\pm0.26$ & $ 9.99\pm0.15$ \\
 8828-0140-1  & $ 333.4$ & $ -49.4$ & 6000 & 3.50 & $-0.75\pm0.08$ & $1.12\pm0.48$ & $ 9.85\pm0.21$ \\
 4977-1226-1  & $ 343.6$ & $  52.9$ & 5033 & 2.55 & $-1.08\pm0.17$ & $1.11\pm0.28$ & $ 9.85\pm0.35$ \\
 8401-0199-1  & $ 346.6$ & $ -27.1$ & 4625 & 1.39 & $-1.21\pm0.17$ & $0.97\pm0.26$ & $            $ \\
 8389-1156-1  & $ 352.1$ & $ -26.4$ & 4713 & 0.97 & $-1.78\pm0.17$ & $1.21\pm0.25$ & $ 9.99\pm0.10$ \\
 \hline
 \end{tabular}\end{center}
 {\small (*)~This star cannot be found by its Tycho number in the SIMBAD electronic database,
 but it has an alternative designation, RAVE J051701.4$-$661815. }
 \end{table}
%%%%%%%%%%%%%%%%%%%%%%%%%%%%%%%%%%%%%%%%%
%%%%%%%%%%%%%%%%%%%%%%%%%%%%%%%%%%%%%%%%% Fig 2
 \begin{figure}[t] {\begin{center}
 \includegraphics[width=160mm]{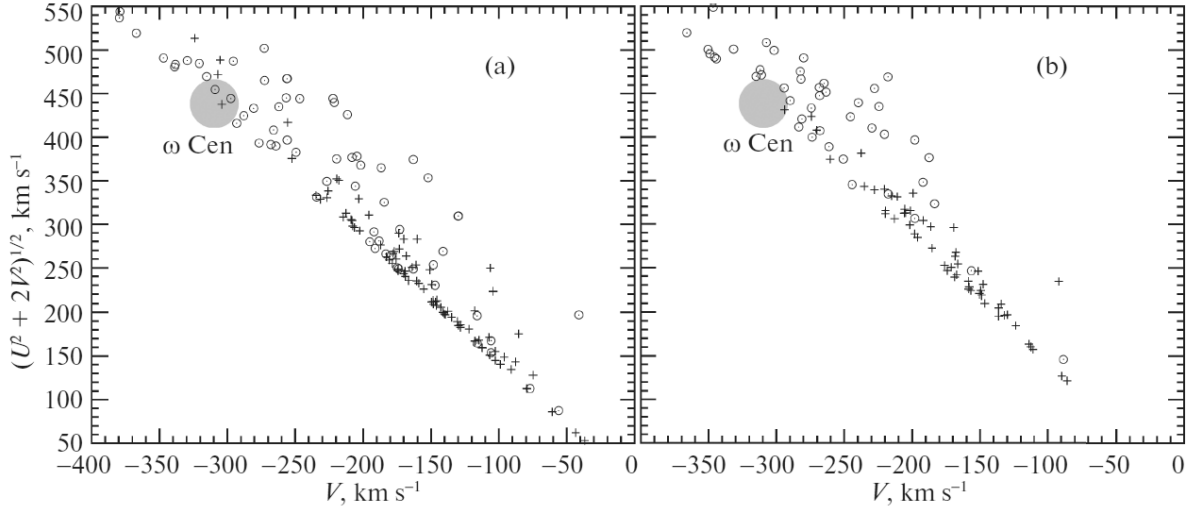}
 \caption{
 Velocity distribution for the low-metallicity stars of nine groups
from Helmi et al. (2017) in the $V,\sqrt{U^2+2V^2}$ plane. The
open circles mark the stars from the VelHel-1, 2, 3, 4, 5, 8, and
9 groups; the crosses mark the stars from the VelHel-6 and
VelHel-7 groups. The big gray circle indicates the position of the
globular cluster $\omega$~Cen. The velocities are given relative
to the Sun: the stellar velocities calculated using the
trigonometric parallaxes (a) and the photometric distances (b).
 }
 \label{f-3-Helmi-all} \end{center} } \end{figure}
%%%%%%%%%%%%%%%%%%%%%%%%%%%%%%%%%%%%%%%%%%%%%%%%%%
%%%%%%%%%%%%%%%%%%%%%%%%%%%%%%%%%%%%%%%%%%%%%%%%%% Fig 3
 \begin{figure}[p] {\begin{center}
 \includegraphics[width=80mm]{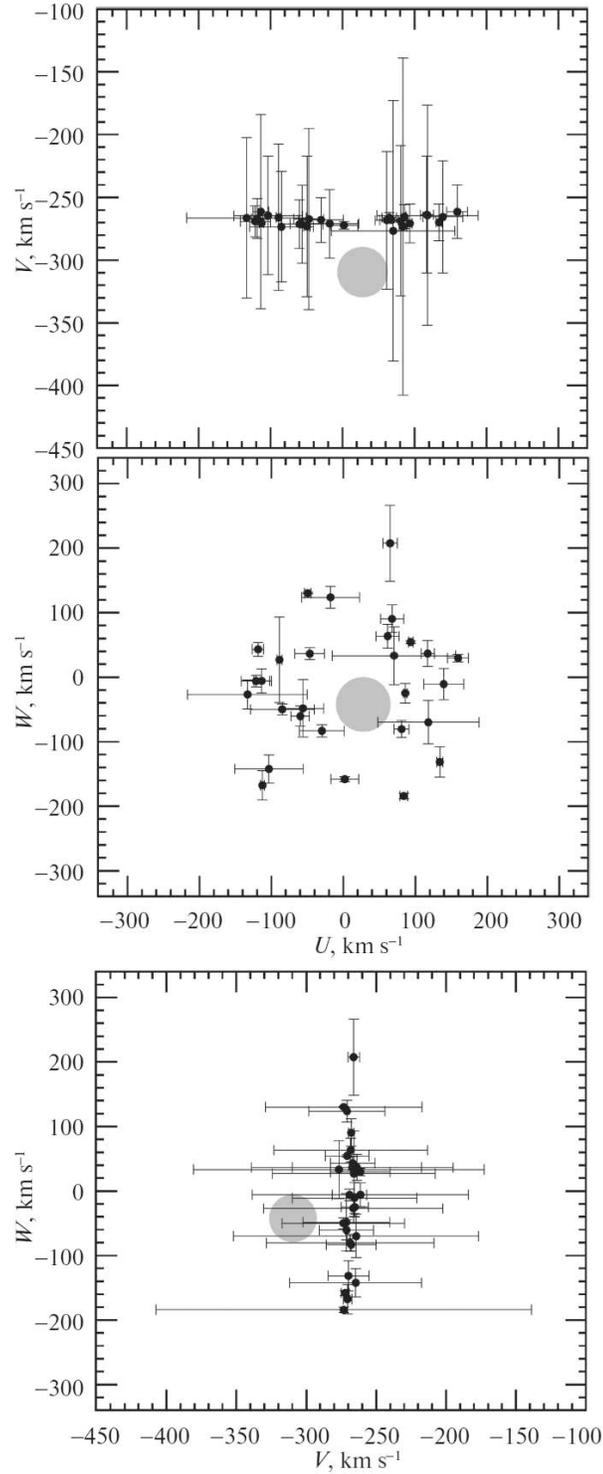}
 \caption{
Stellar velocity distributions for the BB17-3 stream in the $UV$,
$UW$ and $VW$ planes. The gray circle indicates the position of
the globular cluster $\omega$~Cen. The velocities are given
relative to the Sun.
 }
 \label{f-4} \end{center} } \end{figure}
%%%%%%%%%%%%%%%%%%%%%%%%%%%%%%%%%%%%%%%%%
%%%%%%%%%%%%%%%%%%%%%%%%%%%%%%%%%%%%%%%%% Fig 4
 \begin{figure}[t] {\begin{center}
 \includegraphics[width=140mm]{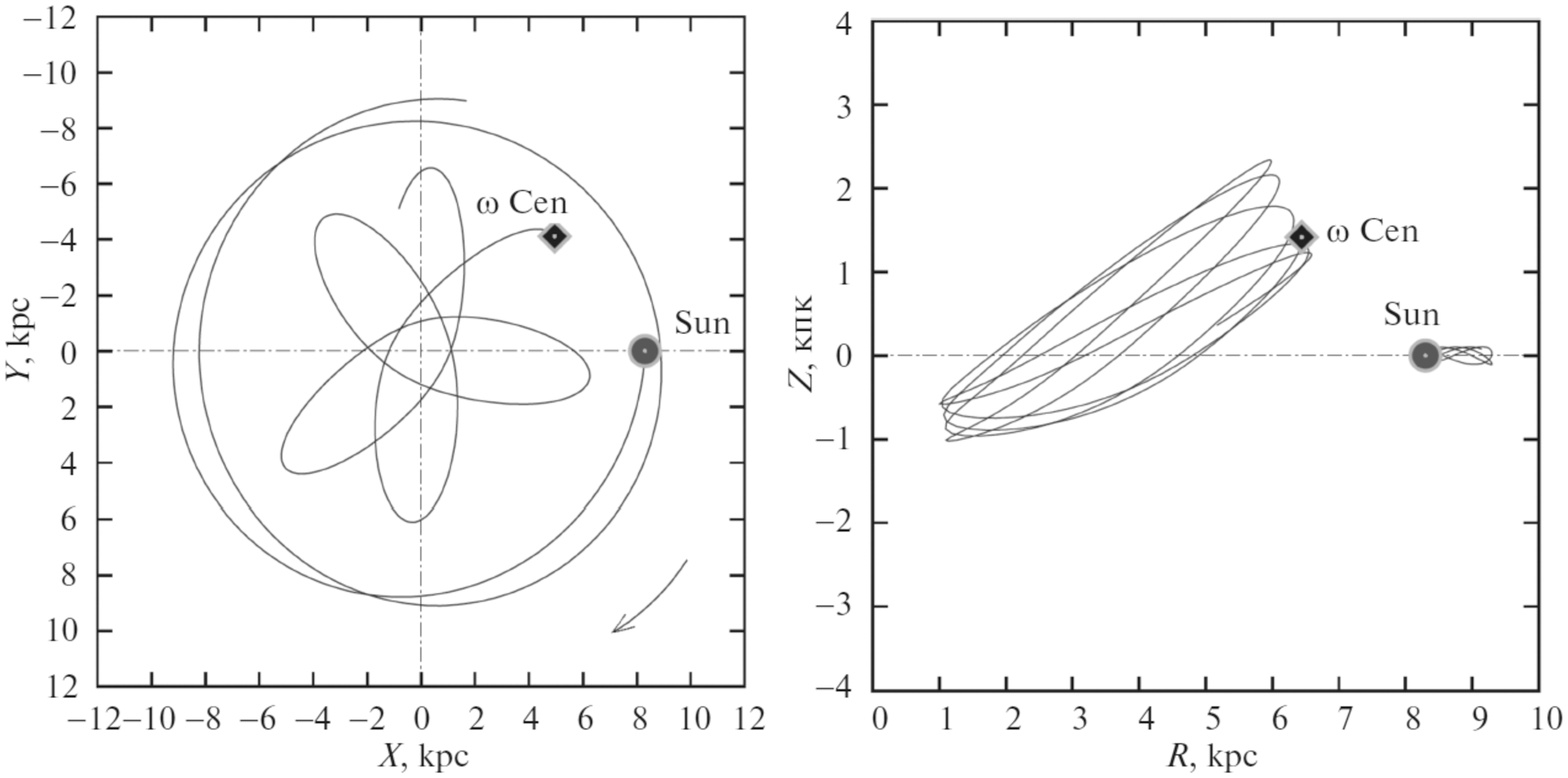}
 \caption{
The Galactic orbits of the Sun and the globular cluster
$\omega$~Cen in projection onto the Galactic $XY$ plane (a) and
the associated $RZ$ plane (b). The orbits were constructed in an
interval of 400 Myr into the future. The arrow on the left panel
indicates the direction of Galactic rotation.
 }
 \label{f-5} \end{center} } \end{figure}
%%%%%%%%%%%%%%%%%%%%%%%%%%%%%%%%%%%%%%%%

 \subsection*{RESULTS}
To identify the low-power clumps in the region of high space
velocities, we apply the approach proposed by Arifyanto and Fuchs
(2006). For this purpose, we consider the distribution of stellar
velocities in the $V,\sqrt{U^2+2V^2}$ plane. Using this approach,
Klement et al. (2008) discovered the KFR08 stream, while Bobylev
et al. (2010) found new bright members of this stream.

Figure 1 presents the wavelet map of the $V,\sqrt{U^2+2V^2}$
velocity plane constructed using the same stars as those in the
previous step. The three BB17-1, BB17-2, and BB17-3 clumps now
manifest themselves especially clearly, while the BB17-3 clump
appears as a completely isolated peak. Interestingly, in Fig. 5
from Bobylev et al. (2010) constructed using Hipparcos data we can
see low-power clumps of stars near the BB17-2 ($V=-220$~km
s$^{-1}$, $\sqrt{U^2+2V^2}=330$~km s$^{-1}$) and BB17-3
($V=-270$~km s$^{-1}$, $\sqrt{U^2+2V^2}=390$~km s$^{-1}$) peaks.

The velocity distribution of low-metallicity stars in the
$V,\sqrt{U^2+2V^2}$ plane for nine groups whose list was taken
from Helmi et al. (2017) is shown in Fig. 2. Here we want to show
that the positions of the stars from the VelHel-1, 2, 3, 4, 5, 8,
and 9 groups are fairly close to the position of the globular
cluster $\omega$~Cen.

Note that Helmi et al. (2017) searched for the groups by analyzing
the integrals of motion, i.e., based on a technique different from
ours. Moreover, they used the photometric distances from the RAVE5
catalogue, because their goal was to analyze the stars
sufficiently far from the Sun, $r>100$~pc. To construct Fig. 2, we
calculated the stellar velocities using both the trigonometric
parallaxes (while ignoring the constraints (4)) and the
photometric distances by taking into account the constraints (4).

The position of the globular cluster $\omega$~Cen in this figure
is marked by the circle whose size exceeds considerably the random
measurement errors of the cluster velocities. Although the radius
of this circle is arbitrary, we wish to emphasize that there is a
noticeable velocity dispersion in the cluster and its debris.
According to the determinations by Sollima et al. (2009), the
stellar velocity dispersion at the center of the cluster
$\omega$~Cen is about 17.2 km s$^{-1}$ and reaches approximately
5.2 km s$^{-1}$, but then remains at a level of 7--8 km s$^{-1}$
for a long time at increasingly large distances from the cluster
center.

According to Helmi et al. (2017), the VelHel-6 and VelHel-7 groups
are not associated with the debris of the globular cluster
$\omega$~Cen. As can be seen from Fig. 2a, most of the crosses are
concentrated near $V=-170$~km s$^{-1}$ and
$\sqrt{U^2+2V^2}=250$~km s$^{-1}$, but rare crosses are also
encountered in the upper part of the diagram, while open circles
are encountered in its lower part. A considerably larger
concentration of open circles to the position of $\omega$~Cen and
a tighter concentration of crosses to the BB17-1 and BB17-2 peaks
(Fig. 1) are clearly seen in Fig. 2b. We can conclude that the
results of our selection of stars based on a fairly simple
technique agree well with the results of applying the method based
on the calculation of the integrals of motion.

We determined the significance level of the revealed peaks by the
method of Monte Carlo simulations as described in Skuljan et al.
(1999). For this purpose, we generated $M=10 000$ random
realizations of $(U,V)$ velocities by taking into account their
measurement errors. We assumed the errors to be additive and to be
distributed according to a normal law with zero mean and a
dispersion equal to the measurement error. For each random
realization we constructed a wavelet map and determined whether
there was a clump in the region of the revealed peaks. The ratio
of the number $N$ of realizations that show the presence of peaks
to the total number $M$ of random realizations gives the
significance level of the peaks. In our case, the significance
level of the new revealed peaks was approximately 91\%. This means
that the detected clumps may be deemed real rather than generated
by random noise with a probability of about 91\%.

The stellar velocity distribution for the BB17-3 stream in the
$UV,$ $UW,$ and $VW$ planes is presented in Fig. 3; the position
of the globular cluster $\omega$~Cen is marked by the circle of an
arbitrary size. Note the $UW$ velocity plane, in which the
velocity dispersions in both coordinates, $\sigma_U$ and
$\sigma_W$, are seen to be equal; however, here we can say nothing
about the dispersion $\sigma_V$ because of the peculiarities of
our selection of stars.

It is well known that an isotropic (Maxwellian) velocity
distribution takes place in spherically symmetric systems, for
example, in globular clusters. At present, the isotropy of the
velocity dispersion has been measured quite reliably for several
globular clusters (Watkins et al. 2015) by invoking the Hubble
Space Telescope observations of stellar proper motions. The mean
ratio of the velocity dispersions in the tangential and radial
directions at the cluster center calculated from the data on 22
globular clusters is $\sigma_t/\sigma_r=0.992\pm0.005$, while for
the globular cluster $\omega$~Cen this ratio is
$\sigma_t/\sigma_r=1.01\pm0.01$.

The velocity dispersion in the debris, of course, will be larger
than that in the parent cluster or galaxy, but the main properties
must be retained. It is also assumed that the stars in the debris
must have a homogeneous chemical composition owing to their common
origin.

The stars of the BB17-3 stream that we selected are listed in the
table. Column 1 gives the numbers according to the Tycho-2
catalogue; columns 2 and 3 give the Galactic coordinates $l,b;$
columns 4--6 give the effective temperature $T_{\hbox {\tiny\rm
eff}},$ gravity $\log~${\it g}, and metallicity [M/H] copied from
the RAVE5 catalogue; the next-to-last column gives the
trigonometric parallax $\pi_{Gaia}$; and the last column gives the
individual age estimate copied from the RAVE4 catalogue
(Kordopatis et al. 2013). Note that the influence of the
well-known Lutz–Kelker bias (Lutz and Kelker 1973) is negligible
at small $\sigma_\pi/\pi$ (10--15\%); otherwise, this bias should
be taken into account (Stepanishchev and Bobylev 2013;
Astraatmadja and Bailer-Jones 2017). As can be seen from the
table, the selected stars mostly have an acceptable mean level of
errors $\sigma_\pi/\pi$, which do not exceed 0.27. In addition,
the entire group is very homogeneous in age.

In fact, we selected the stars based on the boundaries of the
corresponding peak in Fig.~1. For this purpose, we took the stars
in a square with a $V$ side from $-$258~km s$^{-1}$ to $-$278~km
s$^{-1}$ and a $\sqrt{U^2+2V^2}$ side from 380 to 405 km s$^{-1}$.
Our list has no stars common to both the groups of probable
members of the $\omega$~Cen stream proposed by Helmi et al. 2017)
and the list of probable bright members of this stream from
Navarrete et al. (2015). Note that Helmi et al. (2017) selected
the candidates with metallicities [M/H]$\leq-1.5,$ typical for the
lowest-metallicity halo stars. As can be seen from the table, the
metallicities of our selected stars lie in the range [M/H]$\leq$0.
On the whole, this is consistent with the spectroscopic analysis
of stars in the globular cluster $\omega$~Cen by Gratton et al.
(2011), who established the presence of several populations with
different chemical compositions in this cluster. As Villanova et
al. (2014) showed, the subgiants in the globular cluster
$\omega$~Cen form five populations with different metallicities, a
mean age of $\sim$10 Gyr, and a dispersion around this value of
$\sim$2 Gyr.

At the same time, such metallicities and ages are typical for halo
stars (Jofr\'e and Weiss 2011). Therefore, we cannot separate the
halo stars from the stars of the $\omega$~Cen debris in our sample
of BB17-3 clump stars based on their metallicities and ages.

Figure 4 presents the Galactic orbits of the Sun and the globular
cluster $\omega$~Cen constructed in an interval of 400 Myr into
the future. The following input parameters taken from the SIMBAD
electronic astronomical database were used for the cluster
$\omega$~Cen:
 $\alpha=201^\circ.69165,$
 $\delta=-47^\circ.47694,$
 $\mu_\alpha\cos\delta=-6.01\pm0.25$~mas yr$^{-1}$,
 $\mu_\delta=-5.02\pm0.25$~mas yr$^{-1}$,
 $V_r = 238\pm5$~km s$^{-1}$, and
 $d=5.5\pm0.2$~kpc. The peculiar velocity of
the Sun relative to the local standard of rest was assumed to be
$(u_\odot,v_\odot,w_\odot)=(11.1,12.2,7.3)$~km s$^{-1}$, as
determined by Sch\"onrich et al. (2010). To construct the Galactic
orbits of the Sun and the globular cluster $\omega$~Cen, we used
the axisymmetric three-component model of the Galactic potential
with parameters from Bajkova and Bobylev (2016), where it is
designated as model III. The debris of constructing the Galactic
orbits of globular clusters are described in Bobylev and Bajkova
(2017). It can be seen from Fig. 4 that the Sun's motion is
prograde (coincides in direction with the Galactic one), while the
motion of the cluster $\omega$~Cen is retrograde. There is good
agreement of our constructed orbit with the results of other
authors, for example, the orbits from Mizutani et al. (2003) or
Majewski et al. (2012) constructed using other models of the
Galactic gravitational potential.

 \subsection*{DISCUSSION AND CONCLUSIONS}
Let us first say a few words about the main characteristics of the
cluster $\omega$~Cen. Among all of the known globular clusters in
the Galaxy, $\omega$~Cen is one of the biggest. Indeed, it has the
largest size, the greatest oblateness $\varepsilon=0.17$ (Watkins
et al. 2015), the highest luminosity, and the largest mass, whose
estimates lie in the range from $5\times10^6 M_\odot$ (Meylan et
al. 1995) to $2.5\times10^6 M_\odot$ (van de Ven et al. 2006).
Thus, this cluster contains several million stars, but it can also
lose many stars because of the tidal effect of the Galaxy,
especially since the pericenter of the cluster orbit is very close
to the Galactic center (Fig. 4).

(1) An examination of Fig. 4 leads to the question of precisely
how the stars of the $\omega$~Cen debris ended up in the solar
neighborhood. After all, to reach the solar circle, these stars
should traverse approximately 2 kpc in the radial direction.

Note, for example, the results of numerical simulations of the
dynamical evolution of the globular cluster Note, for example, the
results of numerical simulations of the dynamical evolution of the
globular cluster $\omega$~Cen and the formation of its debris
obtained by Ideta and Makino (2004). As these authors showed,
already $\sim$0.44 Gyr after the start time of the dwarf galaxy's
disruption, the width of its debris in the radial direction was
$\sim$4 kpc in some places, while after 0.88 Gyr the debris fills
the solar circle almost completely. In these simulations the
initial mass of the dwarf progenitor galaxy was $1.3\times10^8
M_\odot$ and it was located at an initial distance of $\sim$6 kpc
from the center of the Milky Way.

The results by Meza et al. (2005), who studied the debris of the
origin of a series of debris formed through multiple periodic
encounters of the dwarf progenitor galaxy of the globular cluster
$\omega$~Cen with the Milky Way, are even more interesting. Based
on numerical simulations, these authors showed that during each
encounter the dwarf galaxy left stars with different energies and
angular momenta in the corresponding debris. The debris were
preserved at different Galactocentric distances. Therefore, a
present day observer sees such stars as different kinematic
streams. In these simulations the initial mass of the dwarf galaxy
was $4.6\times10^9 M_\odot$ and it was located at an initial
distance of 140 kpc from the center of the Milky Way and moved in
a highly elliptical orbit.

Mizutani et al. (2003) modeled the kinematic properties of the
tidal disruption of a dwarf galaxy in the attractive field of the
Milky Way. The central part of the dwarf galaxy was assumed to
contain the globular cluster $\omega$~Cen. The motion of the dwarf
galaxy was traced in a long time interval in the past, 1.5--1.9
Gyr, using several models. They confirmed that the motion of the
series of formed debris is retrograde. A strong concentration of
debris stars near the solar circle was found in one of the models
(model 2).

Majewski et al. (2012) pointed that their selected candidates are
concentrated in the fourth Galactic quadrant. Their simulations
showed that most stars from the $\omega$~Cen debris are located
within the solar circle and are concentrated in the fourth
quadrant, near the current position of this cluster. Note that a
similar conclusion follows from an analysis of the Galactic
coordinates for the stars in our table.

(2) All of the identified 28 stars in the BB17-3 clump have very
close ages. The distribution of their metallicity indices
resembles that observed for the stars of the globular cluster
$\omega$~Cen. For example, looking at Fig. 2b, we can conclude
that the stars covering almost the entire upper left corner of the
diagram, i.e., filling the square with a $V$ side from $-250$ to
$-200$~km s$^{-1}$ and a $\sqrt{U^2+2V^2}$ side from 350 to 500~km
s$^{-1}$, belong to the $\omega$~Cen debris. The stars in this
square are distributed quite uniformly. However, as can be seen
from Fig. 1, the 28 stars of the BB17-3 clump form a separate peak
clearly visible by the unaided eye. We can assume with a high
probability that these stars belong to the debris of the cluster
$\omega$~Cen. We also get the impression that they can belong to a
fairly homogeneous ``stream'' in the cluster debris.

(3) The positions of the BB17-1 and BB17-2 peaks agree well with
the positions of the VelHel-6 and VelHel-7 streams (if, for
example, Fig. 1 and Fig. 2b are compared) found by Helmi et al.
(2017). In fact, the presence of two such streams was confirmed in
this paper. At present, no specific cluster or galaxy that could
form debris or streams with such characteristics has been
proposed. Therefore, searching for such objects can become an
interesting task in future.

We can conclude that in this paper the greatest attention is given
to the BB17-3 clump. We showed that it is most closely linked with
the debris of the globular cluster $\omega$~Cen. In this clump we
identified 28 stars having a low velocity dispersion with respect
to the center of their distribution in the $V,\sqrt{U^2+2V^2}$
plane. The list of BB17-3 clump members does not overlap with the
list of members of the VelHel-1, 2, 3, 4, 5, 8, and 9 groups
described by Helmi et al. (2017) and also associated with the
debris of the globular cluster $\omega$~Cen. All these stars have
very close individual age estimates, $\log t\approx10$. The
distribution of metallicity indices in this sample is typical for
the stars of the globular cluster $\omega$~Cen. In our opinion,
the 28 stars of the BB17-3 clump can be characterized as a
homogeneous stream in the extended debris of the cluster
$\omega$~Cen.

 \subsubsection*{ACKNOWLEDGMENTS}
We are grateful to the referee for the useful remarks that
contributed to an improvement of the paper. This work was
supported by the Basic Research Program P--7 of the Presidium of
the Russian Academy of Sciences, the ``Transitional and Explosive
Processes in Astrophysics'' Subprogram. In our study we used the
SIMBAD electronic database.

 \bigskip\medskip{\bf REFERENCES}
{\small

 1. M. I. Arifyanto and B. Fuchs, Astron. Astrophys. 449, 533 (2006).

 2. T. L. Astraatmadja and C. A. L. Bailer-Jones, Astrophys. J. 833, 119 (2017).

 3. A. T. Bajkova and V. V. Bobylev, Astron. Lett. 42, 567 (2016).

 4. V. Belokurov, N. W. Evans, M. J. Irwin, P. C. Hewett, and M. I. Wilkinson,
    Astrophys. J. 637, L29 (2006).

 5. V. V. Bobylev, A. T. Bajkova, and A. A. Myll\"ari, Astron. Lett. 36, 27 (2010).

6. V. V. Bobylev and A. T. Bajkova, Astron. Rep. 61, 551 (2017).

 7. A. G. A. Brown, A. Vallenari, T. Prusti, J. de Bruijne, F. Mignard, R. Drimmel,
    C. Babusiaux, et al. (GAIA Collab.), Astron. Astrophys. 595, 2 (2016).

 8. C. K. Chui, Wavelets: A Mathematical Tool for Signal Analysis
(SIAM, Philadelphia, PA, 1997).

 9. Ya. O. Chumak, A. S. Rastorguev, and S. D. Arset, Astron. Lett. 31, 308 (2005).

 10. Ya. O. Chumak and A. S. Rastorguev, Astron. Lett. 32, 157 (2006a).

 11. Ya. O. Chumak and A. S. Rastorguev, Astron. Lett. 32, 446
(2006b).

 12. W. Dehnen, Astron. J. 115, 2384 (1998).

 13. W. Dehnen, Astron. J. 119, 800 (2000).

 14. O. J. Eggen, Mon. Not. R. Astron. Soc. 118, 65 (1958).

 15. O. J. Eggen, Publ. Astron. Soc. Pacif. 81, 553 (1969).

 16. O. J. Eggen, Publ. Astron. Soc. Pacif. 83, 251 (1971a).

 17. O. J. Eggen, Publ. Astron. Soc. Pacif. 83, 271 (1971b).

 18. O. J. Eggen, Astron. J. 112, 1595 (1996).

 19. R. Fux, Astron. Astrophys. 373, 511 (2001).

 20. R. G. Gratton, C. I. Johnson, S. Lucatello, V. D\'Orazi, and C.
Pilachowski, Astron. Astrophys. 534, 72 (2011).

 21. C. J. Grillmair, K. C. Freeman, M. Irwin, and P. J. Quinn, Astron. J. 109, 2553 (1995).

 22. C. J. Grillmair, and O. Dionatos, Astrophys. J. Lett. 651, L29 (2006).

 23. A. Helmi, Astron. Astrophys. Rev. 15, 145 (2008).

 24. A. Helmi, J. Veljanoski, M. A. Breddels, H. Tian, and L. V. Sales,
     Astron. Astrophys. 598, 58 (2017).

 25. E. Hog, C. Fabricius, V. V. Makarov, U. Bastian, P. Schwekendiek, A. Wicenec,
     S. Urban, T. Corbin, and G. Wycoff, Astron. Astrophys. 355, L27 (2000).

 26. R. Ibata, M. Irwin, G. F. Lewis, and A. Stolte, Astrophys. J. 547, L133 (2001).

 27. M. Ideta and J. Makino, Astrophys. J. 616, L107 (2004).

 28. P. Jofr\'e and A. Weiss, Astron. Astrophys. 533, 59 (2011).

 29. R. Klement, B. Fuchs, and H.-W. Rix, Astrophys. J. 685, 261 (2008).

 30. G. Kordopatis, G. Gilmore, M. Steinmetz, C. Boeche, G. M. Seabroke, A. Siebert,
     T. Zwitter, J. Binney, et al., Astron. J. 146, A134 (2013).

 31. A. Kunder, G. Kordopatis, M. Steinmetz, T. Zwitter, P. McMillan, L. Casagrande,
     H. Enke, J. Wojno, et al., Astron. J. 153, 75 (2017).

 32. S. Leon, G. Meylan, and F. Combes, Astron. Astrophys. 359, 907 (2000).

 33. L. Lindegren, U. Lammers, U. Bastian, J. Hernandez, S. Klioner, D. Hobbs,
     A. Bombrun, D. Michalik, et al., Astron. Astrophys. 595, 4 (2016).

 34. T. E. Lutz and D. H. Kelker, Pub. Astron. Soc. Pacif. 85, 573 (1973).

 35. S. R. Majewski, D. L. Nidever, V. V. Smith, G. J. Damke, W. E.
Kunkel, R. J. Patterson, D. Bizyaev, and A. E. Garc\'ia P\'erez,
Astrophys. J. 747, L37 (2012).

 36. G. Meylan, M. Mayor, A. Duquennoy, and P. Dubath, Astron.
Astrophys. 303, 761 (1995).

 37. A. Meza, J. F. Navarro, M. G. Abadi, and M. Steinmetz, Mon.
Not. R. Astron. Soc. 359, 93 (2005).

 38. A. Mizutani, M. Chiba, and T. Sakamoto, Astrophys. J. 589, L89 (2003).

 39. C. Navarrete, J. Chan\'e, I. Ramirez, A. Meza, G.
Anglada-Escud\'e, and E. Shkolnik, Astrophys. J. 808, 103 (2015).

 40. C. Navarrete, V. Belokurov, and S. E. Koposov, Astrophys. J. 841, 23 (2017).

 41. M. Odenkirchen, E. K. Grebel, C. M. Rockosi, W. Dehnen, R.
Ibata, H.-W. Rix, A. Stolte, et al., Astrophys. J. 548, L165
(2001).

 42. T. Prusti, J.H. J. de Bruijne,A.G. A. Brown, A. Vallenari, C.
Babusiaux, C. A. L. Bailer-Jones, U. Bastian, M. Biermann, et al.
(GAIA Collab.), Astron. Astrophys. 595, 1 (2016).

 43. R. Sch\"onrich, J. Binney, and W. Dehnen, Mon. Not. R. Astron. Soc. 403, 1829 (2010).

 44. J. Skuljan, J. B. Hearnshaw, and P. L. Cottrell, Mon. Not. R. Astron. Soc. 308, 731 (1999).

 45. A. Sollima, M. Bellazzini, R. L. Smart, M. Correnti, E. Pancino, F. R. Ferraro, and
     D. Romano, Mon. Not. R. Astron. Soc. 396, 2183 (2009).

 46. A. Sollima, D. Martinez-Delgado, D. Valls-Gabaud, and J. Penarrubia,
     Astrophys. J. 726, 47 (2011).

 47. M. Steinmetz, T. Zwitter, A. Siebert, F. G. Watson, K. C. Freeman, U. Munari,
     R. Campbell, M. Williams, et al., Astron. J. 132, 1645 (2006).

 48. A. S. Stepanishchev and V. V. Bobylev, Astron. Lett. 39, 185 (2013).

49. G. van de Ven, R. C.E. van den Bosch, E. K. Verolme, and P. T.
de Zeeuw, Astron. Astrophys. 445, 513 (2006).

50. S. Villanova, D. Geisler, R. G.Gratton, and S. Cassisi,
Astrophys. J. 791, 107 (2014).

51. V. V. Vityazev, Wavelet Analysis of Time Series (SPb. Gos.
Univ., St. Petersburg, 2001) [in Russian].

52. L. L. Watkins, R. P. van der Marel, A. Bellini, and J.
Anderson, Astrophys. J. 803, 29 (2015).

 53. The Hipparcos and Tycho Catalogues, ESA SP--1200 (1997).

}
\end{document}